\documentclass[aps,pra,twocolumn,superscriptaddress,nofootinbib,longbibliography,floatfix]{revtex4-2}

\usepackage{graphicx}
\usepackage{amsmath,amssymb}

\begin{document}

\title{Vacuum amplification across a programmable bosonic Bogoliubov stability boundary in a trapped-ion chain}

\author{Lu Qi}
\affiliation{Duke Quantum Center, Duke University, Durham, NC 27701, USA}
\affiliation{Department of Electrical and Computer Engineering, Duke University, Durham, NC 27708, USA}

\author{Duanyang Wang}
\affiliation{Duke Quantum Center, Duke University, Durham, NC 27701, USA}
\affiliation{Department of Physics, Duke University, Durham, NC 27708, USA}

\author{Kenneth R. Brown}
\email{kenneth.r.brown@duke.edu}
\affiliation{Duke Quantum Center, Duke University, Durham, NC 27701, USA}
\affiliation{Department of Electrical and Computer Engineering, Duke University, Durham, NC 27708, USA}
\affiliation{Department of Physics, Duke University, Durham, NC 27708, USA}
\affiliation{Department of Chemistry, Duke University, Durham, NC 27708, USA}

\begin{abstract}
A quantum oscillator always carries vacuum fluctuations, and modulating its parameters converts them into correlated pairs of real quanta. This vacuum amplification underlies squeezed light and Hawking radiation. In a closed quadratic system, pair creation competes with the energy splitting of the modes, and growth turns exponential only once pairing overcomes the splitting. Here we realize this competition in two motional modes of a trapped-ion chain.
A two-mode-squeezing drive creates the pairs, and a beam-splitter coupling splits the motion into two coupled modes at different frequencies. The two rates are tuned and calibrated independently. Starting near the motional ground state, we cross the stability boundary in two ways. First, with the drive at the sum of the two bare-mode frequencies, the phonon number stays bounded until the pairing rate exceeds the splitting and grows exponentially beyond it. Second, tuning the drive into resonance with one coupled mode lowers the threshold to zero, and only that mode is amplified into a squeezed vacuum. Separate control of the two interactions makes the threshold for vacuum amplification programmable. This control can be used to simulate particle creation and quantum phase transitions, and to operate parametrically driven bosonic devices close to their stability limit.
\end{abstract}

\maketitle

\section{Introduction}\label{sec:intro}

The quantum vacuum is not inert. Modulating the parameters of a field or oscillator amplifies its vacuum fluctuations and converts them into correlated pairs of real quanta. This mechanism underlies Hawking radiation \cite{hawking1974,hawking1975}, cosmological particle production \cite{parker1968}, and the dynamical Casimir effect \cite{wilson2011}. The same process produces squeezing. The reduced quadrature noise improves precision measurements \cite{tse2019}, and its two-mode form distributes continuous-variable entanglement \cite{metzner2023,ockeloen2018}. What limits such amplification depends on the system. In an optical parametric oscillator the limit is cavity loss, and amplification sets in where gain exceeds loss. In a coherent quadratic system the limit is the mode spectrum. Pair creation competes with the energy splitting of the modes, and this competition determines whether the response stays bounded.

\begin{figure*}[t]
\centering
\includegraphics[width=\linewidth]{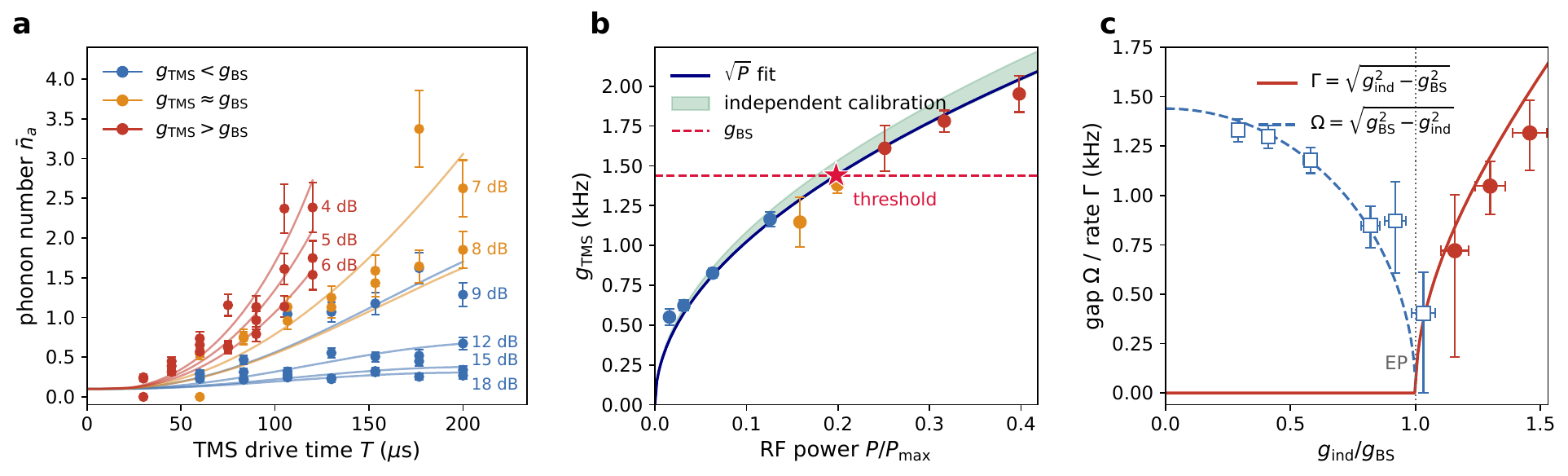}
\caption{\textbf{Crossing the exceptional point at $\Delta=0$.} (a) Bare axial occupation $\bar n_a$ versus TMS drive time at $g_{\mathrm{BS}}=1.44\pm0.05$~kHz, for attenuations 18--4~dB. Colours group the powers below (blue), near (orange) and above (red) threshold. Error bars are uncertainties of the thermal sideband-flop fits. Solid lines are fits of the closed two-mode dynamics of Eq.~\eqref{eq:H} with $g_{\mathrm{TMS}}$ as the only free parameter. The dynamics is propagated through the applied pulse envelope from the post-cooling occupation $\bar n_0=0.10$. (b) The fitted coupling scales as $g_{\mathrm{TMS}}=(3.24\pm0.05)\sqrt{P/P_{\mathrm{max}}}$~kHz and reaches $g_{\mathrm{BS}}$ at $7.0$~dB (star), which locates the threshold. The shaded band is the independent decoupled-point calibration, $(3.33\pm0.11)\sqrt{P/P_{\mathrm{max}}}$~kHz, corrected for the zero-point amplitudes. Error bars are fit uncertainties of $g_{\mathrm{TMS}}$. (c) Bogoliubov gap $\Omega=|\omega_\pm|$ below threshold (open squares) and instability rate $\Gamma=\Gamma_\pm$ above it (filled circles), from one-parameter fits of the traces in (a) at the swap-measured $g_{\mathrm{BS}}$. The horizontal coordinate $g_{\mathrm{ind}}/g_{\mathrm{BS}}$ and the curves $\Omega=\sqrt{g_{\mathrm{BS}}^2-g_{\mathrm{ind}}^2}$ and $\Gamma=\sqrt{g_{\mathrm{ind}}^2-g_{\mathrm{BS}}^2}$ use only the independent calibration of (b) and $g_{\mathrm{BS}}$. The traces in (a) enter only the vertical coordinates. Vertical error bars are fit uncertainties, and horizontal error bars propagate the calibration uncertainty. The gap closes and the rate opens at the exceptional point $g_{\mathrm{ind}}=g_{\mathrm{BS}}$ (dotted line).}
\label{fig:thresh}
\end{figure*}

In its minimal form the competition involves two modes, and its quantum theory was worked out more than sixty years ago in the early treatments of parametric amplification \cite{louisell1961,mollow1967}.
Several experiments have realized parts of this two-mode problem. Spinor condensates amplified vacuum fluctuations \cite{klempt2010}, followed the motion away from the unstable fixed point of a quantum pendulum \cite{gerving2012}, and studied quenches across a continuous quantum phase transition \cite{hoang2016}. In these experiments the spin-exchange energy is set by the condensate, and no number-conserving coupling mixes the two created modes. Trapped ions realized single-mode phonon-pair creation \cite{wittemer2019} and phase-sensitive two-mode squeezing with SU(1,1) interferometry \cite{metzner2023}, without simultaneous number-conserving mode mixing and pair creation. Sonic-horizon experiments have reported spontaneous pair correlations and thermal, stationary spontaneous emission consistent with Hawking physics \cite{steinhauer2016,steinhauer2019,steinhauer2021}, with the amplified mode structure set by the flow profile rather than by separately tunable couplings. Driven--dissipative platforms do combine beam-splitter and two-mode-squeezing couplings \cite{ockeloen2018,slim2024}. However, their modes are strongly damped, and amplification sets in only where gain exceeds loss.
To our knowledge, no experiment has crossed a boundary set by a beam-splitter splitting against pairing starting from the quantum vacuum.

Here we realize this two-mode problem with two motional modes of a trapped-ion chain. A static beam-splitter (BS) interaction splits the modes into two coupled modes by a voltage-tunable amount, and a driven two-mode-squeezing (TMS) interaction creates pairs within them. Each parameter of the two-mode Hamiltonian is set by its own hardware control and calibrated independently, so that the stability boundary can be placed within the accessible range and crossed from the motional ground state.
With the drive tuned to the symmetric point midway between the two coupled-mode resonances, raising its power increases the pairing rate and turns a bounded oscillation of the phonon number into exponential growth. The transition occurs where the pairing rate equals the splitting.
Tuning the drive to the resonance of one coupled mode lowers this threshold to zero, and that mode alone is amplified into a squeezed vacuum, about ten times beyond the bound that holds at the symmetric point.
Separate control of the two rates makes the threshold for vacuum amplification programmable, so the same system can simulate particle creation and quantum phase transitions. Because the threshold is fixed by two coherent rates and not by loss, it also remains the operating limit of parametrically driven bosonic devices as their loss is engineered away.

\section{Results}\label{sec:results}

\subsection{Programmable two-mode Bogoliubov dynamics}\label{subsec:programmable}

\begin{figure*}[t]
\centering
\includegraphics[width=0.72\linewidth]{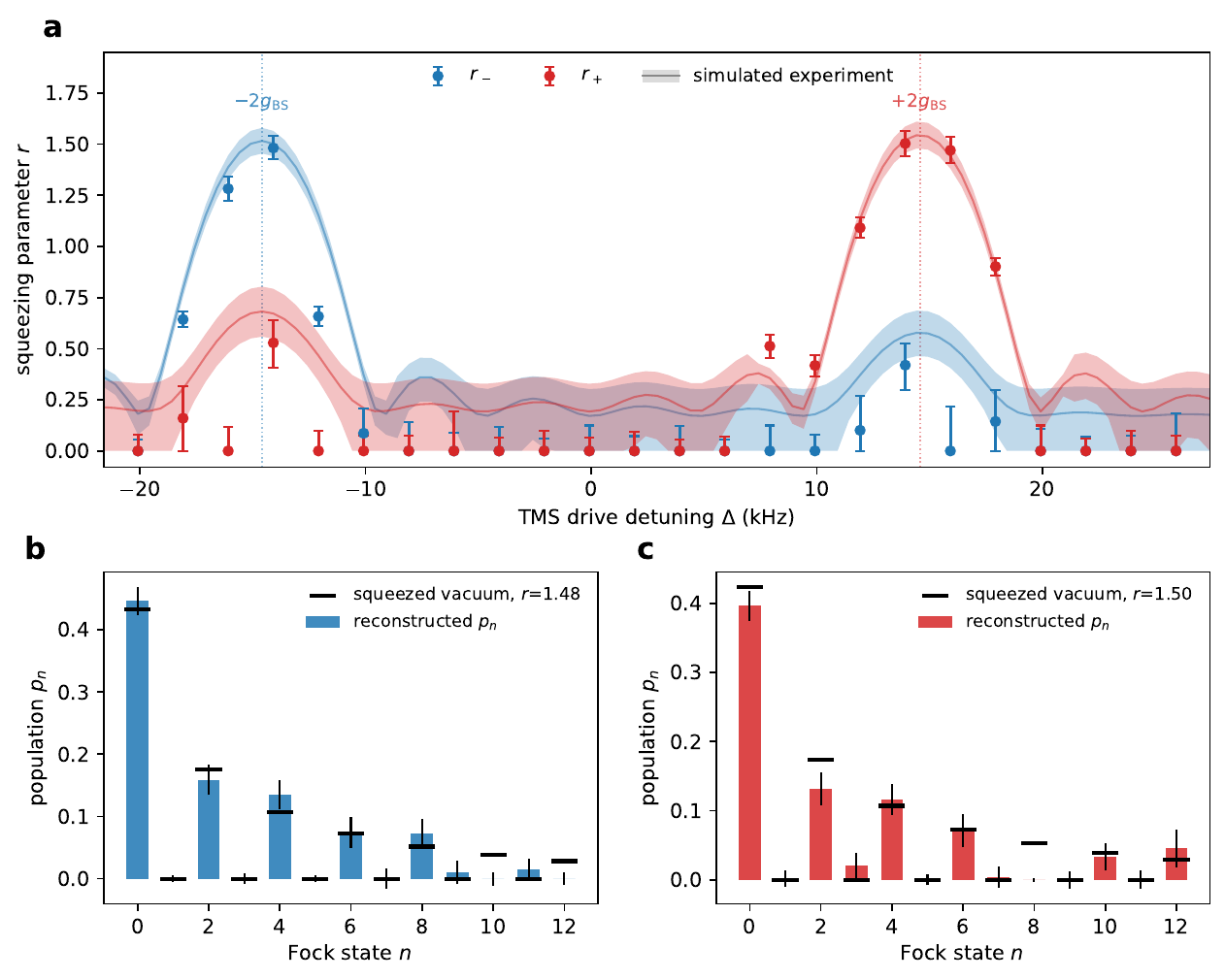}
\caption{\textbf{Detuning-selected amplification of the two coupled modes.} (a) Squeezing $r_\pm$ of each coupled mode versus TMS drive frequency after a $220~\mu$s drive pulse at fixed power. Each mode is read out on its own sideband. Each point is a squeezed-vacuum fit to the readout flop, with profile-likelihood ($\Delta\chi^2=1$) error bars.
Shaded bands show a simulated experiment. At each drive frequency, the two-mode state produced by the detuned parametric drive is integrated over the pulse envelope, with the coupling fixed by the on-resonance peaks alone ($g_{\mathrm{TMS}}=1.13$ and $1.15$~kHz). This state is read out through the exact sideband dynamics, including the off-resonant response of the sideband of the other mode. The resulting flop is analyzed with the same squeezed-vacuum fit as the data (Methods). The band is the profile-likelihood ($\Delta\chi^2=1$) interval of that fit at each point's number of repetitions. The simulation assumes the ideal spectral sidelobes of the Tukey pulse and treats the leaked flop as fully coherent, so away from the two resonances the band is a conservative upper estimate of the apparent squeezing.
(b,c) Fock populations of $c_-$ (b) and $c_+$ (c) at their own peaks, fitted to the sideband flops without assuming a population model (Methods). Black ticks mark the ideal squeezed-vacuum ladder at the fitted $r$. The odd-$n$ population is $P_{\mathrm{odd}}=0.053\pm0.041$ ($c_-$) and $0.123\pm0.050$ ($c_+$), against $0.37\pm0.04$ for simulated flops of the best-fit thermal state. This suppression of odd phonon numbers is the parity fingerprint of pairwise phonon generation. Error bars in (b,c) are bootstrap intervals over the measurement shots.}
\label{fig:asym}
\end{figure*}

We use the axial and a radial out-of-phase modes $a$ and $b$ of a two-ion chain (Methods), coupled by a static BS interaction and driven by a TMS interaction. In the frame rotating at half the TMS drive frequency ($\hbar=1$),
\begin{equation}
\begin{split}
H ={}& \frac{\delta}{2}\left(a^\dagger a - b^\dagger b\right) - \frac{\Delta}{2}\left(a^\dagger a + b^\dagger b\right)\\
&+ g_{\mathrm{BS}}\left(a^\dagger b + a b^\dagger\right) + g_{\mathrm{TMS}}\left(a^\dagger b^\dagger + a b\right),
\end{split}
\label{eq:H}
\end{equation}
where $\delta=\omega_a-\omega_b$ is the bare-mode detuning and $\Delta$ is the detuning of the TMS drive from the sum of the two bare-mode frequencies, $\omega_a+\omega_b$ (conventions and calibration in Methods). At $\delta=0$, the BS coupling dresses the oscillators into $c_\pm=(a\pm b)/\sqrt{2}$ with opposite frequency shifts $\pm g_{\mathrm{BS}}$, and the TMS term creates pairs within each coupled mode rather than across the pair. Equation~\eqref{eq:H} then separates into two independent single-mode parametric problems, $H_\pm=d_\pm c_\pm^\dagger c_\pm+(\lambda_\pm/2)(c_\pm^{\dagger 2}+c_\pm^2)$, with signed effective detuning and pairing amplitude
\begin{equation}
d_\pm=\pm g_{\mathrm{BS}}-\frac{\Delta}{2},\qquad
\lambda_\pm=\pm g_{\mathrm{TMS}},
\label{eq:signed}
\end{equation}
in which $d_\pm$ is the frequency of the coupled mode $c_\pm$ measured from half the drive frequency, and $\lambda_\pm$ is the amplitude with which pairs are created inside it.
When $|\lambda_\pm|>|d_\pm|$, the coupled mode nearest resonance becomes unstable and sets the predicted critical coupling
\begin{equation}
g_c(\Delta)=\min\left(|d_+|,|d_-|\right)=\left|g_{\mathrm{BS}}-\frac{|\Delta|}{2}\right|.
\label{eq:boundary}
\end{equation}
For $|\lambda_\pm|<|d_\pm|$ the Bogoliubov transformation $c_\pm=\cosh s_\pm\,b_\pm-\sinh s_\pm\,b_\pm^\dagger$, with $\tanh2s_\pm=\lambda_\pm/d_\pm$, brings each coupled mode to $H_\pm=\omega_\pm b_\pm^\dagger b_\pm$ up to a constant, with $\omega_\pm=\operatorname{sgn}(d_\pm)\sqrt{d_\pm^2-\lambda_\pm^2}$. The dynamical matrix of that coupled mode has the eigenvalues $\pm\omega_\pm$, one for the Bogoliubov mode $b_\pm$ and one for its conjugate. At $g_{\mathrm{TMS}}=g_c$ the two eigenvalues of the coupled mode nearest resonance meet at zero, $s_\pm$ diverges, and their eigenvectors become parallel. The dynamical matrix of that mode can then no longer be diagonalized, and the boundary is an exceptional point. Below $g_c$ the eigenvalues are real and pair creation is coherently reversed. For a square pulse applied to the vacuum, the occupation then oscillates within a bound,
\begin{equation}
\begin{gathered}
\bar n_\pm(t)=\frac{\lambda_\pm^2}{d_\pm^2-\lambda_\pm^2}
\sin^2\!\left(\sqrt{d_\pm^2-\lambda_\pm^2}\,t\right),\\
r_{\max}=\operatorname{artanh}\!\left(\frac{|\lambda_\pm|}{|d_\pm|}\right),
\end{gathered}
\label{eq:stablebound}
\end{equation}
so the squeezing never exceeds $r_{\max}$, whatever the drive time. At $g_c$ the occupation grows quadratically in time, $\bar n=\lambda_\pm^2t^2$. Above $g_c$ the eigenvalues become $\pm i\Gamma_\pm$, with the instability rate $\Gamma_\pm=\sqrt{\lambda_\pm^2-d_\pm^2}$, and the vacuum is amplified exponentially, $\bar n\propto\sinh^2(\Gamma_\pm t)$.

Equation~\eqref{eq:H} carries four parameters, and each is set by its own hardware control with its own calibration (Methods). A compensation voltage sets the splitting through $g_{\mathrm{BS}}$, the drive power sets the pairing magnitude $g_{\mathrm{TMS}}$, the drive frequency sets $\Delta$, and the radial RF amplitude sets the bare-mode detuning $\delta$. The measurements below are made at $\delta=0$. Equation~\eqref{eq:boundary} traces the exceptional points as detuning shifts the coupled modes relative to the drive, and the two measurements below cross this boundary along orthogonal directions, in pairing strength and in detuning.

\subsection{Crossing the exceptional point}\label{subsec:threshold}

We first fix $\Delta=0$, where the threshold $g_c=g_{\mathrm{BS}}$ is highest, and raise the pairing strength through the splitting. We set $g_{\mathrm{BS}}=1.44\pm0.05$~kHz, vary the TMS drive power, and measure the bare axial occupation after different drive times (Fig.~\ref{fig:thresh}). At $\Delta=0$ the two coupled modes have equal $|d_\pm|=g_{\mathrm{BS}}$ and $|\lambda_\pm|=g_{\mathrm{TMS}}$ and evolve independently, so $\bar n_a$ follows the occupation of either coupled mode. The response is bounded at low power and grows exponentially at high power. We fit each trace with the closed two-mode dynamics of Eq.~\eqref{eq:H}, propagated through the applied pulse envelope from the post-cooling occupation $\bar n_0=0.10$, with $g_{\mathrm{TMS}}$ as the only free parameter. The fitted coupling scales as the square root of the drive power, $g_{\mathrm{TMS}}=(3.24\pm0.05)\sqrt{P/P_{\mathrm{max}}}$~kHz (Fig.~\ref{fig:thresh}b), and reaches $g_{\mathrm{BS}}$ at an attenuation of $7.0$~dB. An independent calibration at the decoupled point gives $(3.33\pm0.11)\sqrt{P/P_{\mathrm{max}}}$~kHz after correction for the frequency dependence of the zero-point amplitudes (Methods).
The two calibrations agree to within $3\%$.

At $\Delta=0$ the pairing pulls the two Bogoliubov frequencies in from $\pm g_{\mathrm{BS}}$ toward zero.
Physically, the splitting and the pairing compete. Below threshold, the splitting makes the phase of the pair correlation $\langle c_\pm^2\rangle$ rotate faster than the drive amplifies it, so successive pairs partly cancel and the phonon number stays bounded. Above threshold, the amplification outpaces this rotation and the phonon number grows exponentially. At threshold the two rates are equal, and the dynamical matrix reaches an exceptional point.
The gap $\Omega=|\omega_\pm|$ and $\Gamma=\Gamma_\pm$ in Fig.~\ref{fig:thresh}c come from one-parameter pulse-envelope fits of the growth traces at the swap-measured splitting. The gap decreases from $1.33\pm0.06$~kHz at the weakest drive to $0.40^{+0.21}_{-0.40}$~kHz at $7$~dB, and the instability rate reaches $1.32^{+0.17}_{-0.19}$~kHz at the strongest drive. At $\Delta=0$, $c_+$ and $c_-$ reach their exceptional points together, so the threshold is where both gaps close.

The threshold is determined by the relative strengths of the coherent couplings.
The axial and mixed modes heat at $0.8\pm0.1$ and $4.2\pm1.8$~quanta\,s$^{-1}$, and the swap coherence time of $1.43\pm0.10$~ms is more than six times the longest drive (Methods). As a result, the dynamics is effectively coherent on the experimental timescale.

\subsection{Frequency selection of the amplified mode}\label{subsec:detuning}

Equation~\eqref{eq:boundary} predicts that detuning lowers the threshold, which reaches zero when the drive is resonant with a coupled mode, $\Delta=\pm2g_{\mathrm{BS}}$. We test this at $g_{\mathrm{BS}}=7.40\pm0.01$~kHz, where a weak sideband probe resolves the two coupled modes as separate lines and reads out each of them individually (Methods). We prepare both modes near the motional ground state, apply the drive for $220~\mu$s at fixed power, and scan $\Delta$ across both resonances (Fig.~\ref{fig:asym}a). Each coupled mode is amplified only when the drive is tuned to its own resonance, and the two resonances lie $28\pm2$~kHz apart, consistent with $4g_{\mathrm{BS}}$. Dedicated measurements at the two peaks give $r_-=1.48\pm0.06$ and $r_+=1.50\pm0.06$. The two values agree, as expected from the sub-percent difference of the two zero-point amplitudes.

To test the number statistics without assuming a population model, we fit each sideband flop at the two peaks as a sum of single-Fock-state Rabi oscillations, with the populations $p_n$ as free parameters \cite{jia2022} (Fig.~\ref{fig:asym}b,c and Methods). Both modes show the even ladder of a squeezed vacuum, with odd-number weights of $0.05\pm0.04$ for $c_-$ and $0.12\pm0.05$ for $c_+$. These populations identify the squeezed vacuum in the number basis. A thermal model fitted directly to the same flops is rejected in both modes (Methods). The phase coherence of the trapped-ion two-mode-squeezing interaction was established separately by SU(1,1) interferometry, where the anti-squeezed quadrature is resolved directly \cite{metzner2023}.

A coupled mode driven at $d_\pm=0$ has $\Gamma_\pm=|\lambda_\pm|$ and follows the pulse-area law $r=g_{\mathrm{TMS}}T_{\mathrm{eff}}$.
With the $208~\mu$s effective pulse area, the two peaks give $g_{\mathrm{TMS}}=1.14\pm0.05$~kHz, in agreement with $1.18\pm0.04$~kHz from the independent power calibration (Methods). The coupling ratio $g_{\mathrm{TMS}}/g_{\mathrm{BS}}=0.154\pm0.007$ is far below the threshold crossed in Fig.~\ref{fig:thresh}. At this ratio Eq.~\eqref{eq:stablebound} limits the squeezing at the symmetric point to $r_{\max}=0.155\pm0.007$, about ten times below the resonant values. For the other coupled mode, which stays $2g_{\mathrm{BS}}$ from its own resonance, the limit is $0.077$.
Away from the resonance of the amplified mode, both cross readouts are consistent with this limit. Near that resonance, the readout of the other coupled mode gives a fitted squeezing of up to $0.5$. This residual is spectral leakage of the readout from the amplified mode, not excitation of the other mode. A simulation of the leakage with no free parameters gives a comparable value near the peak centre (shaded bands in Fig.~\ref{fig:asym}a and Methods).

Physically, the splitting places the two coupled modes at different frequencies, so the drive can be tuned to amplify one of them. Tuning the drive to $\Delta=\pm2g_{\mathrm{BS}}$ sets $d_\pm=0$ for that mode. The phase of that mode no longer rotates relative to the drive, and the mode is amplified at the full rate $|\lambda_\pm|$. The other mode stays detuned by $2g_{\mathrm{BS}}$, its phase rotates many times relative to the drive during the pulse, and it undergoes only the bounded transient. The sign of the detuning therefore sets which coupled mode is amplified.

\section{Discussion}\label{sec:discussion}

Both measurements cross the boundary of Eq.~\eqref{eq:boundary} at $\delta=0$, where the bare modes are degenerate. Detuning the bare modes opens a third instability window at $\Delta=0$, where the threshold is highest for degenerate modes. A nonzero $\delta$ raises the splitting to $\sqrt{4g_{\mathrm{BS}}^2+\delta^2}$ and divides the pairing between two channels.
Pairing within each coupled mode keeps the amplitude $g_{\mathrm{TMS}}\sin2\theta$, while pairing between coupled modes takes $g_{\mathrm{TMS}}\cos2\theta$, with $\tan2\theta=2g_{\mathrm{BS}}/\delta$.
The frequencies of the two coupled modes, measured from half the drive frequency, are $\pm\tfrac12\sqrt{4g_{\mathrm{BS}}^2+\delta^2}$.
Because these frequencies sum to zero, at $\Delta=0$ the drive pairs the two coupled modes resonantly, and nothing opposes this cross-mode amplification.
In the lossless model any nonzero $\delta$ therefore removes the threshold at $\Delta=0$.
For $g_{\mathrm{TMS}}\ll g_{\mathrm{BS}}$ the resulting growth rate is approximately $g_{\mathrm{TMS}}|\cos2\theta|$, which is small when $|\delta|\ll g_{\mathrm{BS}}$.
The calibration of $g_{\mathrm{TMS}}$ uses the opposite limit, $|\delta|\gg g_{\mathrm{BS}}$, where all of the pairing is cross-mode and grows without threshold (Methods).

In a frame where both coupled-mode frequencies are positive, $\Delta<-2g_{\mathrm{BS}}$, Eq.~\eqref{eq:H} is the quadratic normal-phase expansion of an anisotropic Dicke model \cite{emary2003}, and the threshold coincides with its superradiant instability. Ramping the pairing slowly on this side of the boundary prepares the normal-phase ground state, so the present controls can follow the closing gap and the growing ground-state squeezing toward the critical point. Replacing one oscillator by the Ca$^+$ internal qubit would add the spin nonlinearity that stabilizes the superradiant phase beyond the critical coupling \cite{hwang2015}. A spin-dependent force with independent red- and blue-sideband amplitudes provides this coupling, as in the quantum Rabi model realized with a single trapped ion \cite{cai2021}.

The boundary also matters for quantum hardware. For parametrically driven mediators \cite{ge2019,burd2021}, the power scan and the detuning scan sample a two-dimensional stability margin. Once pairing exceeds the splitting, or the drive reaches a coupled-mode resonance, the mediator runs away and can no longer be adiabatically eliminated. Circuit-QED and optomechanical devices combine BS and TMS interactions for directional amplification and reservoir engineering \cite{abdo2013,sliwa2015,lecocq2017,ockeloen2018}, with damping built into the design. The coherent margin measured here is the limit that remains as that damping is engineered away.

The same interaction pair has further uses that require phase-sensitive control and readout. It is the single bond of the bosonic Kitaev chain, with the beam-splitter coupling as hopping and pairing as the anomalous term \cite{mcdonald2018,slim2024}. Extending it along a chain of motional modes would realize that model in a trapped-ion crystal. With the quadratures $q=(a+a^\dagger)/\sqrt2$ and $p=-i(a-a^\dagger)/\sqrt2$ of each mode, the two interactions read $g_{\mathrm{BS}}(a^\dagger b+ab^\dagger)+g_{\mathrm{TMS}}(a^\dagger b^\dagger+ab)=(g_{\mathrm{BS}}+g_{\mathrm{TMS}})\,q_aq_b+(g_{\mathrm{BS}}-g_{\mathrm{TMS}})\,p_ap_b$. At the exceptional point, $g_{\mathrm{TMS}}=g_{\mathrm{BS}}$ with $\delta=\Delta=0$, the $p_ap_b$ term vanishes and only the quadrature product $2g_{\mathrm{BS}}\,q_aq_b$ remains. This generator conserves both $q$ quadratures and shears the $p$ quadratures linearly in time, which is the origin of the quadratic growth at threshold. It is also the continuous-variable controlled-phase generator that entangles grid-state qubits \cite{gottesman2001,fluhmann2019}.
The pairing term is also the generator of horizon pair creation, the two-mode squeezing that converts vacuum fluctuations into Hawking quanta and their partners. Past threshold, the same generator realizes in two modes the collision of positive- and negative-norm modes that drives the black-hole-laser instability \cite{corley1999,coutant2010}, with the pairing and the splitting set directly rather than by a background flow.

\begin{acknowledgments}
This work is supported by a collaboration between the U.S. DOE and other Agencies. This material is based upon work supported by the U.S. Department of Energy, Office of Science, National Quantum Information Science Research Centers, Quantum Systems Accelerator. Additional support is acknowledged from the Army Research Office (W911NF-21-1-0346 and W911NF-26-1-A063) and the ARO Spectator Qubit MURI (W911NF-18-1-0218).
\end{acknowledgments}

\section*{Data and code availability}

The data supporting the findings of this study will be deposited in a public repository upon publication. The analysis code used to extract squeezing parameters, perform the squeezed-vacuum/thermal model selection, and fit the instability rates will be made publicly available upon publication.

\section*{Competing interests}

The authors declare no competing interests.

\appendix

\section{Methods}\label{sec:methods}

\subsection{Ion chain, trap, and mode preparation}
The experiments are performed on a Ca$^+$--CaOH$^+$ two-ion chain in a segmented blade trap held in an ultra-high-vacuum chamber at a pressure of $\sim5\times10^{-11}$~Torr. The chain is prepared by loading two Ca$^+$ ions and leaking O$_2$ gas into the chamber until one of them reacts to form CaO$^+$. The CaO$^+$ then reacts with the background gas to form CaOH$^+$. A magnetic field of 6.2~G along the trap axis provides the quantization axis. Doppler cooling and state detection use a linearly polarized 397~nm beam and an 866~nm repumping beam on Ca$^+$. Electromagnetically-induced-transparency (EIT) cooling uses a linearly polarized and a $\sigma$-polarized 397~nm beam. Coherent operations and sideband readout on $S_{1/2}\leftrightarrow D_{5/2}$ use a 729~nm laser, and an 854~nm laser quenches the $D_{5/2}$ population.

\begin{figure*}[t]
\centering
\includegraphics[width=\linewidth]{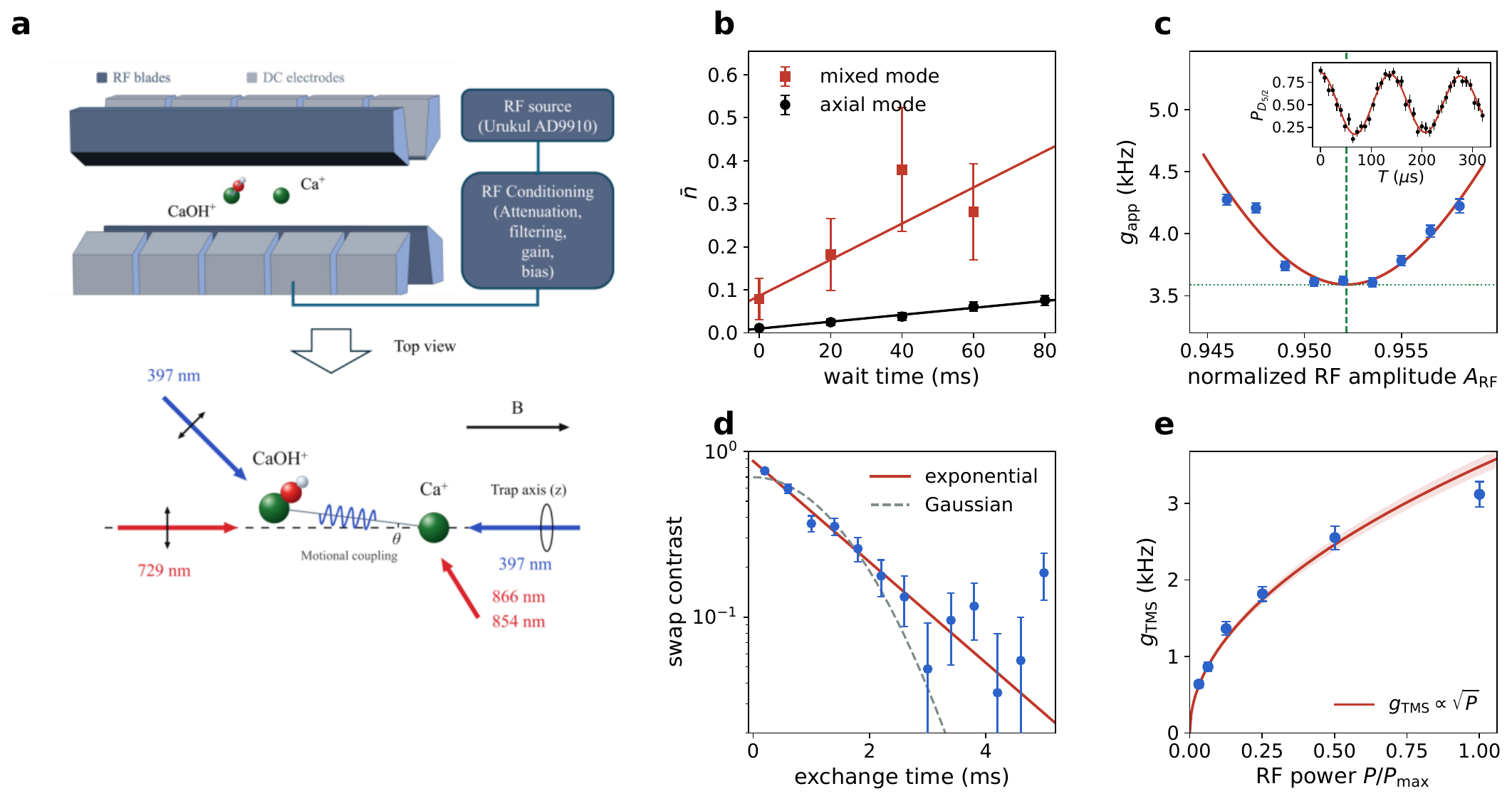}
\caption{\textbf{Experimental system and independent calibrations.} (a) Ca$^+$--CaOH$^+$ two-ion chain in the blade trap and the RF chain that synthesizes the mode-coupling drives. The chain runs from an Urukul AD9910 direct digital synthesizer through a step attenuator (FW-6+), amplifier (ZHL-3A+), low-pass filter (BLP-1.9), variable attenuator (VAT-3+) and bias-tee (ZFBT-R2GW-FGT+) to a trap electrode. Laser beams are labelled by wavelength and polarization (Methods). (b) Motional heating from sideband-ratio thermometry versus wait time. The axial mode heats at $\dot{\bar n}=0.8\pm0.1$~quanta\,s$^{-1}$ (black) and the mixed mode at $4.2\pm1.8$~quanta\,s$^{-1}$ (red). Over the drive time, $\dot{\bar n}\,T\ll1$. (c) Location of $\delta=0$ and $g_{\mathrm{BS}}$ at the working point. The coupling $g_{\mathrm{app}}=\tfrac12\sqrt{\delta^2+4g_{\mathrm{BS}}^2}$ from phonon-swap Ramsey is plotted versus the normalized radial RF amplitude $A_{\mathrm{RF}}$. The hyperbola minimum fixes the degeneracy at $A_{\mathrm{RF}}^0=0.9522\pm0.0001$ and $g_{\mathrm{BS}}=3.59\pm0.02$~kHz. Inset, the swap-Ramsey trace at the minimum, where $g_{\mathrm{app}}\simeq g_{\mathrm{BS}}$. (d) Phonon-swap coherence. The windowed swap contrast decays exponentially with $\tau=1.43\pm0.10$~ms (exponential strongly preferred over Gaussian). (e) $g_{\mathrm{TMS}}$ calibration at the decoupled point, following $g_{\mathrm{TMS}}=(3.48\pm0.11)\sqrt{P/P_{\mathrm{max}}}$~kHz.}
\label{fig:methods}
\end{figure*}

The trap provides a single-Ca$^+$ radial secular frequency tunable over 0.8--1.2~MHz and an axial frequency tunable over 100--450~kHz. A DC electric field lifts the degeneracy of the two radial directions, splitting the single-ion radial frequencies by about 30~kHz, so each radial mode of the chain appears as a pair. We operate at a radial potential of 1.1~MHz and an axial potential of 0.45~MHz. Because of the mass difference between Ca$^+$ and CaOH$^+$, the radial modes are strongly localized. The mass-weighted participation ratio is 98\% Ca$^+$ and 2\% CaOH$^+$ for the higher radial mode near 1.06~MHz, and the reverse for the molecule-dominated radial mode near 0.72~MHz. The axial modes are shared, with 67\% Ca$^+$ and 33\% CaOH$^+$ for the OP mode near 0.72~MHz and 33\% Ca$^+$ and 67\% CaOH$^+$ for the IP mode near 0.41~MHz. This choice of potentials places the axial OP mode close to the molecule-dominated radial modes, and the exact degeneracy is then located from the phonon swap described below.

Because the molecule-dominated radial modes carry almost no Ca$^+$ participation, they cannot be cooled efficiently through the Ca$^+$. We cool them by mode mixing \cite{wang_prep}. The preparation sequence is as follows. First, the chain is Doppler cooled on all modes. Second, the Ca$^+$-dominated radial modes and the axial IP mode are cooled with EIT cooling. Third, the axial OP mode is tuned into resonance with one radial mode of the CaOH$^+$, and the mixed pair is cooled to $\bar n\approx0.1$ with EIT cooling followed by resolved sideband cooling. The axial OP mode is then ramped adiabatically into resonance with the other CaOH$^+$ radial mode \cite{qi2023}, and this pair is cooled to $\bar n\approx0.1$ in the same way. This second radial mode and the axial OP mode are the modes $b$ and $a$ of Eq.~\eqref{eq:H} on which the analogue is realized.

After preparation, we measure the heating rate by inserting a variable delay before sideband thermometry (Fig.~\ref{fig:methods}b). For both modes the heating accumulated over a drive time of a few hundred microseconds is far below one quantum, so the coherence budget is set by the phonon-swap coherence time discussed below.

\subsection{Calibration of $g_{\mathrm{BS}}$ and swap coherence}

The beam-splitter interaction is generated by the same mechanism used to cool the molecule-dominated modes. A DC bias applied in the radial plane displaces the two ions by different amounts from the RF null and tilts the ion chain away from the trap axis. The tilt introduces cross terms such as $x\cdot z$ and $y\cdot z$ in the trapping potential, which couple the axial and radial motion. When the two mode frequencies are brought into resonance, the coupling takes the beam-splitter form of Eq.~\eqref{eq:H} \cite{qi2023,fabrikant2024}. The coupling rate $g_{\mathrm{BS}}$ is set by the magnitude of the radial-plane bias. Measurements versus the compensation-electrode voltage give a linear local slope $\mathrm{d}g_{\mathrm{BS}}/\mathrm{d}V=0.37\pm0.01$~kHz\,V$^{-1}$.

To calibrate $g_{\mathrm{BS}}$, we prepare a single phonon in mode $a$ with a 729~nm sideband pulse on the Ca$^+$ and record the population swap between the two modes. The swap population oscillates at $2g_{\mathrm{BS}}$, and a fit to the oscillation yields $g_{\mathrm{BS}}$ (Fig.~\ref{fig:methods}c, inset). A finite mode detuning biases the coupling rate upward, since the fit returns $g_{\mathrm{app}}=\tfrac12\sqrt{\delta^2+4g_{\mathrm{BS}}^2}$, which equals $g_{\mathrm{BS}}$ only at $\delta=0$. We therefore locate $\delta=0$ as the minimum of the phonon swap frequency versus the normalized radial RF amplitude $A_{\mathrm{RF}}$, which tunes the radial mode into degeneracy with the axial OP mode (Fig.~\ref{fig:methods}c). We re-establish this point before every run. We measure the coherence of the swap by extending it to many periods and fitting the decay of its contrast. The phonon-swap coherence time is $\tau=1.43\pm0.10$~ms, more than six times the longest drive (Fig.~\ref{fig:methods}d).

\subsection{Calibration of $g_{\mathrm{TMS}}$}

\begin{figure*}[t]
\centering
\includegraphics[width=0.8\linewidth]{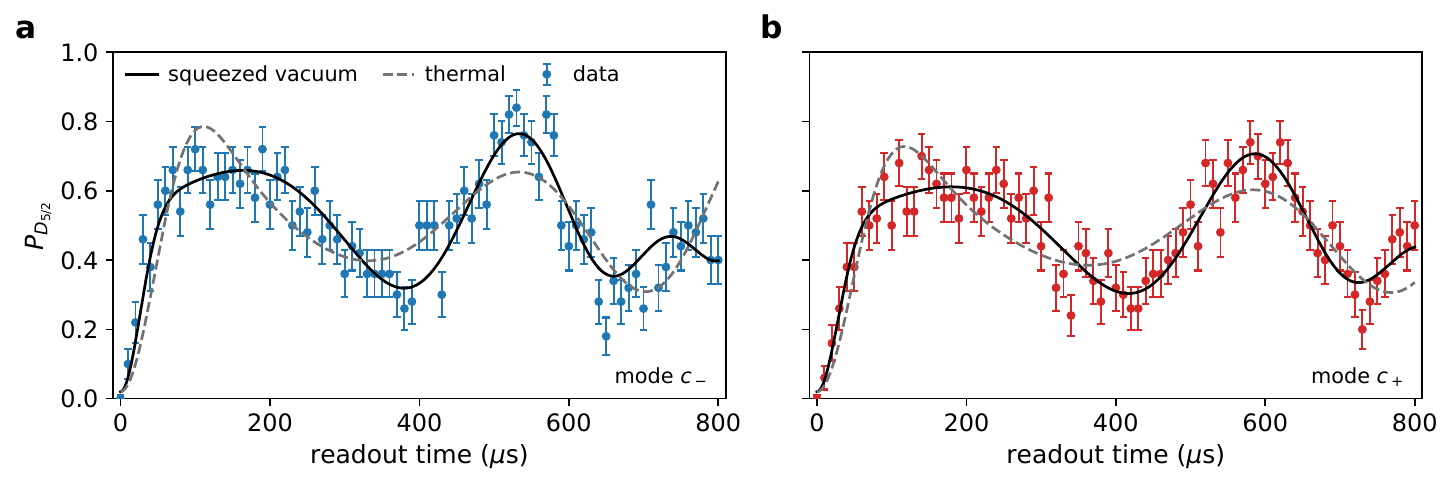}
\caption{\textbf{Mode-resolved readout flops at the two peaks.} Blue-sideband readout
flops of $c_-$ (a) and $c_+$ (b) at their own peaks in Fig.~\ref{fig:asym}a, the traces
behind the Fock populations of Fig.~\ref{fig:asym}b,c. Solid lines are squeezed-vacuum fits
($r=1.48$ and $1.50$). Dashed lines are the best thermal fits, worse by $\Delta\chi^2=117$ ($c_-$)
and $95$ ($c_+$) over $78$ degrees of freedom. Both fits use the calibrated readout Rabi
frequency and coherence envelope (text).}
\label{fig:flops}
\end{figure*}

The TMS coupling is calibrated with the mode mixing effectively switched off. We detune the axial OP mode far from mode $b$, $|\delta|\gg g_{\mathrm{BS}}$, so that the bare modes decouple, and drive the TMS interaction at $\omega_a+\omega_b$. In the ideal quadratic model each bare mode then evolves into the thermal marginal of a two-mode squeezed vacuum, with mean occupation $\bar n=\sinh^2(g_{\mathrm{TMS}}T)$ after a drive time $T$. We measure $\bar n_a(T)$ with sideband thermometry on the bare axial mode, fitting the flops with a thermal population model, and extract $g_{\mathrm{TMS}}$ from the growth of $\bar n_a$. All TMS drives in this work are Tukey-windowed pulses rather than square pulses. Their smooth edges span many periods of the MHz drive and suppress the spectral sidelobes of a sharp edge, which would otherwise excite the motional modes off resonance. At the decoupled point the accumulated squeezing depends only on the pulse area, so the calibration pulses, with $20~\mu$s edges, are analyzed exactly as square pulses of effective duration $T_{\mathrm{eff}}=T-20~\mu$s, which enters the growth law above. The main-text measurements use $12~\mu$s edges, giving the $208~\mu$s effective area of the Fig.~\ref{fig:asym} pulse, and the threshold fits of Fig.~\ref{fig:thresh} propagate the full pulse envelope. Repeating the calibration versus drive power gives $g_{\mathrm{TMS}}\propto\sqrt{P}$, as expected for a coupling linear in the drive amplitude. At the decoupled calibration point, $g_{\mathrm{TMS}}=(3.48\pm0.11)\sqrt{P/P_{\mathrm{max}}}$~kHz (Fig.~\ref{fig:methods}e). The threshold scan uses the coupling fitted \emph{in situ} at its own working point, $g_{\mathrm{TMS}}=(3.24\pm0.05)\sqrt{P/P_{\mathrm{max}}}$~kHz (Fig.~\ref{fig:thresh}b). The two prefactors are related by the frequency dependence of the zero-point amplitudes, $g_{\mathrm{TMS}}\propto1/\sqrt{\omega_a\omega_b}$, between the two mode configurations. Scaling the decoupled-point calibration by this factor gives $g_{\mathrm{TMS}}(P)=(3.33\pm0.11)\sqrt{P/P_{\mathrm{max}}}$~kHz, in $3\%$ agreement with the in-situ fit and independent of the threshold data.
The mode-resolved scan of Fig.~\ref{fig:asym} uses a fixed drive attenuation of $9$~dB on the second chain, $P/P_{\mathrm{max}}=12.6\%$. The calibration predicts $g_{\mathrm{TMS}}=1.18\pm0.04$~kHz. The resonant pulse-area law $r=g_{\mathrm{TMS}}T_{\mathrm{eff}}$, with the $208~\mu$s effective pulse area and the two measured peaks $r_-=1.48\pm0.06$ and $r_+=1.50\pm0.06$, gives $1.14\pm0.05$~kHz, in agreement with the prediction.
Figure~\ref{fig:thresh}a,b use the in-situ fit. In Fig.~\ref{fig:thresh}c the horizontal coordinate and the theory curves use only the independent calibration and the swap-measured $g_{\mathrm{BS}}$. The vertical coordinates are fit from the growth traces at the same splitting, so the two estimates of $g_{\mathrm{TMS}}$ are independent.

\subsection{Branch discrimination below and above threshold}

Over a short drive window the bounded and the growing solutions of the envelope model can resemble each other. To confirm the branch assignments in Fig.~\ref{fig:thresh}c, we refit every trace twice, constrained to $g_{\mathrm{TMS}}\leq g_{\mathrm{BS}}$ and to $g_{\mathrm{TMS}}\geq g_{\mathrm{BS}}$. The two branches separate through their amplitude and late-time curvature. Below threshold the bounded response develops downward curvature within the measured window, which the growing branch cannot reproduce. Above threshold the accelerating growth excludes the bounded branch. The disfavoured branch is rejected with $\Delta\chi^2\geq15$ at every attenuation except the 7~dB crossing, where $\Delta\chi^2=6$. The two branches are expected to merge at the threshold, and the model-inferred gap there, $\Omega=0.40^{+0.21}_{-0.40}$~kHz, is consistent with zero.
At high occupation the sideband-flop readout loses sensitivity to the absolute phonon number. The threshold fits therefore use the calibrated dynamic-range mask $\bar n<5$. The high-power threshold traces reach this readout ceiling, and points with $\bar n\geq5$ are excluded.

\subsection{Mode-resolved readout and number statistics}

The coupled modes are read out on their blue sidebands of the 729~nm transition, which the weak spectroscopic probe resolves as two lines $2g_{\mathrm{BS}}$ apart. To address a single coupled mode the readout is attenuated to sideband Rabi frequencies of $\Omega_0=2.97\pm0.03$~kHz for $c_-$ and $2.67\pm0.03$~kHz for $c_+$. These are calibrated from drive-off flops of the ground-state-cooled modes, which show full-contrast single-mode oscillations ($t_\pi=168$ and $187~\mu$s) with no beat at the frequency of the other coupled mode. The same flops fix the exponential coherence envelope used in all fits below.

At the two peaks, the squeezed-vacuum and thermal fits use the same Fock-ladder flop with the calibrated envelope. The three free parameters of each are the state parameter ($r$ or $\bar n$), the contrast and the readout Rabi frequency, which is floated within $\pm6\%$ of the calibration. The fitted Rabi frequencies run $3$--$6\%$ below the cold calibration, with the same sign as the Debye--Waller reduction of the sideband Rabi frequencies at high phonon number. The thermal model fits the same flops worse by $\Delta\chi^2=\chi^2_{\mathrm{th}}-\chi^2_{\mathrm{sq}}=117$ for $c_-$ and $95$ for $c_+$ over $78$ degrees of freedom (Fig.~\ref{fig:flops}). We also fit the same flops with free Fock populations, without assuming a population model \cite{jia2022}. Level $n$ oscillates at $\Omega_0\sqrt{n+1}$, and the populations are constrained to $p_n\geq0$ and $\sum_np_n=1$. This fit gives the odd-number weights of Fig.~\ref{fig:asym}b,c, $0.05\pm0.04$ for $c_-$ and $0.12\pm0.05$ for $c_+$. The same fit applied to synthetic flops returns $0.37\pm0.04$ for the best-fit thermal state and $0.15\pm0.05$ for a squeezed vacuum at the fitted $r$. The measured odd weights therefore match squeezed vacuum and exclude thermal statistics in both modes.

Each readout also responds off resonantly to the sideband of the other coupled mode, $2g_{\mathrm{BS}}\approx15$~kHz away. For a mode near the motional ground state this response is negligible, as the drive-off flops show. When the other coupled mode is amplified to $r\approx1.5$, its populated levels reach Fock-enhanced Rabi frequencies $\Omega_n=\Omega_0\sqrt{n+1}\approx6$--$8$~kHz, comparable to the sideband separation. Part of its flop then leaks into the readout. The squeezed-vacuum model fits the leaked signal as a squeezing of the read mode. To estimate this contribution, we simulate the readout flop for the two-mode state produced by the drive at each frequency, using the calibrated Rabi frequency, envelope and sideband separation. We fit the simulated flop with the same squeezed-vacuum model as the data. The simulation has no free parameters. Near the centre of each peak it gives a squeezing comparable to the measured residual of up to $0.5$ (shaded bands in Fig.~\ref{fig:asym}a). Away from the resonance of the amplified mode, both cross readouts are consistent with the bounded-transient limit of $0.077$ derived below.

\subsection{Evaluation of the bounded-transient limit}

Equation~\eqref{eq:stablebound} is the bounded-branch solution for a square pulse. At the symmetric point, where each coupled mode has effective detuning magnitude $g_{\mathrm{BS}}$, it gives $r_{\max}=\operatorname{artanh}(0.154)=0.155\pm0.007$ for the coupling ratio of the detuning scan. For the other coupled mode, where the effective detuning is $2g_{\mathrm{BS}}$, it gives $\operatorname{artanh}(0.154/2)=0.077$. With the $12~\mu$s-edge Tukey envelope of the experimental pulse, the same bounded-branch dynamics gives at most $0.151$ and $0.069$ for drive times from $30$ to $400~\mu$s, slightly below the square-pulse values.

\bibliography{Bogoliubov}

\end{document}